\documentclass[reprint,aps,prl,amsmath,amssymb,superscriptaddress, preprintnumbers]{revtex4-2}
 
\usepackage[utf8]{inputenc}
\usepackage[T1]{fontenc}

\usepackage{amsmath}
\usepackage{mathrsfs}
\usepackage{txfonts}
\usepackage{mathtools}
\usepackage{braket}
\usepackage{tensor}
\usepackage{xcolor}
\usepackage{graphicx}
\usepackage{booktabs} 
\usepackage{float}

\usepackage{braket}
\usepackage[normalem]{ulem}

\newcommand{\nene}{$^{20}$Ne+$^{20}$Ne}
\newcommand{\oooo}{$^{16}$O+$^{16}$O}

\usepackage{hyperref}
\usepackage{cleveref}

\newcommand{\beq}{\begin{equation}}
\newcommand{\eeq}{\end{equation}}
\newcommand{\beqn}{\begin{eqnarray}}
\newcommand{\eeqn}{\end{eqnarray}}
\newcommand{\bsub}{\begin{subequations}}
\newcommand{\esub}{\end{subequations}}
\newcommand{\bpm}{\begin{pmatrix}}
\newcommand{\epm}{\end{pmatrix}}

\usepackage[dvipsnames]{xcolor}

\begin{document} 

\title{\textit{Ab initio} anatomy of quadrupole correlations in $^{16}$O and $^{20}$Ne}
     
\author{Chenrong Ding}
\affiliation{School of Physics and Astronomy, Sun Yat-sen University, Zhuhai 519082, P.R. China}    
\affiliation{Guangdong Provincial Key Laboratory of Quantum Metrology and Sensing, Sun Yat-Sen University, Zhuhai 519082, China }

\author{Benjamin Bally}
\email{benjamin.bally@tu-darmstadt.de}
\affiliation{Technische Universit\"at Darmstadt, Department of Physics, 64289 Darmstadt, Germany}

\author{Stavros Bofos}
\affiliation{CEA, DES, IRESNE, DER, SPRC, LEPh, 13108 Saint-Paul-lez-Durance, France}

\author{Thomas Duguet}
\email{thomas.duguet@cea.fr}
\affiliation{Université Paris-Saclay – CEA – IRFU, 91191 Gif-sur-Yvette, France}  

\author{Yi Li}
\affiliation{School of Physics and Astronomy, Sun Yat-sen University, Zhuhai 519082, P.R. China}    
\affiliation{Guangdong Provincial Key Laboratory of Quantum Metrology and Sensing, Sun Yat-Sen University, Zhuhai 519082, China } 

\author{Jiangming Yao}
\email{yaojm8@sysu.edu.cn}
\affiliation{School of Physics and Astronomy, Sun Yat-sen University, Zhuhai 519082, P.R. China}    
\affiliation{Guangdong Provincial Key Laboratory of Quantum Metrology and Sensing, Sun Yat-Sen University, Zhuhai 519082, China } 
\affiliation{Yukawa Institute for Theoretical Physics, Kyoto University, Kyoto, 606-8502, Japan}  
    
\date{\today}

\begin{abstract}

Azimuthal hadronic flow measured in ultra-relativistic ion--ion collisions provides a new means of imaging multipole correlations in the ground state of atomic nuclei. Early interpretations largely relied on a classical-rotor picture, in which the measured mean-square elliptic flow is directly related to an intrinsic quadrupole deformation. Atomic nuclei, however, contain additional many-body correlations generated by the Pauli exclusion principle, collective shape fluctuations and non-collective dynamical processes, whose impact on this correspondence has not yet been elucidated. Here, we resolve this issue through an \textit{ab initio} analysis of $^{16}$O and $^{20}$Ne based on chiral nuclear interactions, combining the in-medium similarity renormalization group with the quantum-number-projected generator coordinate method. By successively isolating antisymmetrization, collective rotational and vibrational, and non-collective dynamical correlations, we determine, for the first time, how each component contributes to the mean-square quadrupole eccentricity.
We uncover an unexpected compensation among these distinct correlation mechanisms: despite sizable individual contributions, the squared \textit{effective} quadrupole deformation inferred from the elliptic flow remains close to the squared \textit{intrinsic} deformation of the nucleus. This result provides a microscopic explanation for the surprising success of the classical-rotor approximation and establishes a quantitative foundation for interpreting \oooo~and \nene~collision data recently collected at the Large Hadron Collider.

\end{abstract}

\preprint{}

\maketitle

 
\paragraph{Introduction.} The atomic nucleus is a finite quantum many-body system in which nucleons self-organize as a result of their collective and non-collective dynamics induced by the strong and electromagnetic interactions between them. This organization manifests itself indirectly, and differently, through various nuclear observables. For a majority of nuclei, angular correlations among nucleons leave characteristic fingerprints in $k$-nucleon ($k>1$) correlation functions that have been interpreted phenomenologically in terms of {\it intrinsic} nuclear deformation parameters associated with the classical picture of nuclear shape rotations and vibrations. 

Ultra-relativistic heavy-ion collisions  (URHICs) dedicated to the study of the  quark-gluon plasma (QGP), a state of nuclear matter similar to the first moments after the Big Bang, have recently emerged as a powerful tool to directly image multi-nucleon correlations in nuclear ground states~\cite{Giacalone2018a,Giacalone2021a,Summerfield2021a,Bally2022a,Zhang2022a,Ryssens2023a,STAR2024a,Giacalone2025a,Giacalone2025b,STAR:2025elk}.  The instantaneous character of high-energy collisions (of the order of $10^{-26}$s) is such that both nuclei remain in their respective ground state throughout the reaction. Following the formation and cooling down of the QGP, anisotropies in the azimuthal flow of hadrons detected in the final state of ultra-central collisions result, to leading order, from spatial (e.g., angular) correlations among nucleons in the collided nuclei~\cite{Niemi:2016}. In particular, multipolar anisotropies in the relative azimuthal angle between pairs of hadrons in symmetric central collisions reflect multipolar moments of the two-nucleon correlation function in the ground state of the collided nuclei~\cite{Blaizot:2025scr,Duguet2025a}.  

As demonstrated recently, it is indeed practically possible to perform such an imaging and extract nuclear structure information in the form of {\it effective} multipolar deformation parameters~\cite{Giacalone:2026fat}. But such a procedure and the resulting interpretation rely on the use of an {\it a priori} given modeling of the collided nuclei. In the first phase of development of this interdisciplinary endeavor, the connection between the anisotropy in the azimuthal hadronic flow and many-body correlations in nuclei was mainly based on the {\it classical rotor} (CR) nuclear model~\cite{Duguet:2025qxi}. While successful at explaining the dominant patterns observed when colliding doubly open-shell nuclei, such a simplistic nuclear structure model leaves unexplained the observation of an anisotropic flow in ultra-central symmetric collisions of doubly closed-shell nuclei beyond quantum fluctuations associated with the finite number of nucleons~\footnote{The present description describes the anisotropic hadronic flow based on pure nucleonic degrees of freedom. While the impact of sub-nucleonic degrees of freedom is expected to be largely subleading, it has to be consistently evaluated in the future.}. 
More generally, a truly quantitative analysis linking the correlated behavior of nucleons in nuclei to final-state observables of ion-ion collisions requires the use of a full-fledged quantum-mechanical framework~\cite{Duguet2025a}.

Building on recent developments~\cite{Bofos:2026huw,Bofos:2026nmg}, we study the quadrupole mean-square eccentricity of the entropy density deposited in ultra-central \oooo~and \nene~collisions, which is proportional to the mean-square elliptic flow~\cite{Niemi:2016,Sousa:2024msh}. Recent measurements at the Large Hadron Collider (LHC) revealed a larger elliptic flow in central \nene~than in \oooo~collisions~\cite{ATLAS:2025nnt,CMS:2025tga,ALICE:2025luc}, in agreement with predictions that attribute this difference to the stronger quadrupole collectivity of $^{20}$Ne~\cite{Giacalone2025a,Giacalone2025b}. A quantitative interpretation, however, requires disentangling the distinct nuclear correlations that generate the initial spatial anisotropy. Having similar masses but remarkably different collective properties, $^{16}$O and $^{20}$Ne represent an especially clean and stringent test case~\cite{Giacalone2025a,CMS:2025tga,ATLAS:2025nnt,Constantin:2025ova}.

In this Letter, the first \textit{ab initio} analysis of how two-body quadrupole correlations affect the mean-square elliptic flow in these systems is presented. By solving the nuclear many-body Schrödinger equation based on a chiral Hamiltonian, we separately quantify collective rotational and vibrational correlations~\cite{Bofos:2026huw}, Pauli correlations~\cite{Bofos:2026nmg}, and non-collective dynamical correlations. Going beyond Ref.~\protect\cite{Blaizot2025a}, we isolate the contribution of each component and assess whether the \textit{effective} quadrupole deformation inferred from URHICs can be identified with the \textit{intrinsic} deformation assumed in the CR model to be the sole source of quadrupole collectivity. An unexpected compensation among the different correlation effects is found to keep the squared effective deformation inferred from the measured mean-square elliptic flow close to the squared intrinsic deformation, thereby providing a microscopic support to the central assumption of the simplified CR picture. 

\paragraph{Theoretical framework.}
The present \textit{ab initio}  theoretical description relies on a nuclear Hamiltonian $H$ derived from chiral effective field theory \cite{Hammer2020a,Epelbaum2020a} that includes two- and three-nucleon interactions. More specifically, the successful EM1.8/2.0~\cite{Hebeler11a} Hamiltonian is expanded using a one-body spherical harmonic oscillator basis (sHO) characterized by the frequency  $\hbar\omega$ and the truncation parameter $e_\mathrm{1max}$~\cite{Hergert20}. Furthermore, three-body matrix elements are further truncated to three-particle states characterized by $e_\mathrm{3max} < 3 e_\mathrm{1max}$

Based on this nuclear Hamiltonian, the many-body Schrödinger equation is solved for ground state of an even-even nucleus, corresponding to the lowest $J=0$ state,
\begin{equation}
    H \ket{\Psi^{J=0}_{n=1}} = E^{0}_1 \ket{\Psi^{0}_1} \, ,
\end{equation}
using the ansatz 
\begin{equation}
\label{ansatz1} 
\ket{\Psi^{0}_1} \equiv \lim_{s\to\infty} | \Psi^{0}_1(s) \rangle  \equiv \lim_{s\to\infty}   U(s) \ket{ \Theta^{0}_1  }_{\text{QRV}}  \, .
\end{equation}

In Eq.~\eqref{ansatz1}, the {\it reference state} $\ket{\Theta^{0}_1}_{\text{QRV}}$ is constructed to efficiently capture collective (long-range) rotational and vibrational correlations. We use the projected generator coordinate method (PGCM) \cite{Hill1953a,Griffin1957a} and consider a linear superposition of symmetry-projected axially deformed mean-field states
\begin{equation}
\begin{split}
 \label{ansatz2}
 \ket{\Theta^{0}_1}_{\text{QRV}} & \equiv \sum_{\beta_{20}} f^{0}_{1,\, \beta_{20}} P^{J=0} P^{\text{N}\text{Z}} \ket{\Phi(\beta_{20})} , \\
 &\equiv \sum_{\beta_{20}} f^{0}_{1,\, \beta_{20}} \ket{\Theta^{0} (\beta_{20})}_{\text{QR}} \,  .
 \end{split}
\end{equation}
Each mean-field state $\ket{\Phi(\beta_{20})}$ is obtained solving constrained Hartree-Fock-Bogoliubov (HFB) equations imposing parity and axial symmetries and such that it carries an {\it intrinsic} axial quadrupole deformation
\begin{equation}
\beta_{20} \equiv \frac{4\pi}{5} \frac{\langle \Phi(\beta_{20}) | Q_{20}  | \Phi(\beta_{20}) \rangle}{\langle  \Phi(\beta_{20}) | r^2 |  \Phi(\beta_{20}) \rangle}\, , 
\label{defdef}
\end{equation}
with $Q_{20} \equiv \sum_{i=1}^{\text{A}} r^2_i Y_{20}(\theta_i,\phi_i)$ and $r^2 \equiv \sum_{i=1}^{\text{A}} r^2_i$, where $r_i$ denotes the nucleon position and $Y_{20}$ a spherical harmonics. The operator $P^J P^{NZ}$ projects the state onto good total angular momentum $J$, neutron number $N$ and proton number $Z$ \cite{Sheikh21a,Bally2021b}. Finally, the weights $f^{0}_{1,\, \beta_{20}}$ are variational parameters obtained by minimizing the energy of $\ket{\Theta^{0}_1}_{\text{QRV}}$.

In Eq.~(\ref{ansatz1}), $U(s)$ represents a continuous unitary transformation depending on a flow parameter $s$ that progressively incorporates non-collective (short-range) correlations on top of $\ket{\Theta^{0}_1}_{\text{QRV}}$. These correlations are typically modeled as incoherent sums of a large number of low-rank individual, i.e., particle-hole, excitations. The transformation is obtained by solving the flow equation defining the in-medium similarity renormalization group (IMSRG) method \cite{Hergert:2016PR}
\begin{equation}
   \frac{d H(s)}{ds} = [\eta(s), H(s)] \, ,
\end{equation} 
with  $H(s)\equiv U^\dagger(s)HU(s)$. The generator of the transformation $\eta(s)$ is meant to be chosen such that the IMSRG evolution, if carried out exactly, makes the reference state the {\it exact} ground state of $H(s)$ as $s\to\infty$. This feature, however, is not guaranteed for multi-reference states such as $\ket{\Theta^{0}_1}_{\text{QRV}}$~\cite{Hergert:2016PR,Hergert:2017PS,Frosini2022c,Duguet:2022zup}. Furthermore, the IMSRG scheme is applied approximately, truncating at the normal-ordered two-body level throughout the flow, i.e., using the IMSRG(2) approximation~\cite{Hergert:2016PR}. To alleviate the computational complexity of the approach, following Ref.\cite{Yao:2020PRL}, the reference state is in fact taken as the statistical {\it ensemble} of the QR state at the spherical point ($\beta_{20}=0$) and the variational optimal QR state, with weights of 80\% and 20\%, respectively. The sensitivity to these weights is gauged in the End Matter (EM).  A similar strategy has been employed with success in previous works~\cite{Yao:2020PRL,Zhou:2024vlt} such that it does not spoil our general conclusions.

Characterizing the physical content of the various states at play, the wave function $\ket{\Theta^{0} (\beta_{20})}_{\text{QR}}$, obtained after symmetry restoration at a fixed value of $\beta_{20}$, describes the quantum rotational motion associated with a single intrinsic shape and corresponds to a {\it quantum rotor} (QR) approximation. The reference state $\ket{\Theta^{0}_1}_{\text{QRV}}$ is obtained by further mixing QR states carrying different intrinsic quadrupole deformations $\beta_{20}$ to incorporate collective shape fluctuations along the axial quadrupole-deformation coordinate. This corresponds to superimposing the vibrational motion on the rotational dynamics such that the resulting state denotes a {\it quantum rotor plus vibrator} (QRV) approximation. Finally, the state $\ket{\Psi^{0}_1}$ additionally includes non-collective dynamics and represents the maximally correlated solution considered in this work.

To partially compensate for the approximations employed during the IMSRG decoupling up to a given value of the flow $s$, an additional PGCM calculation using $H(s)$ is performed. We thus generate a new set of projected mean-field states, labeled QR($s$), carrying an intrinsic deformation $\beta_{20}(s)$, as well as a PGCM state QRV($s$) that is used to define the final approximate ground state in Eq.~\eqref{ansatz1}. The variational nature of the PGCM ensures the relevance of the state QRV($s$) thus generated.

The elliptic anisotropy of the azimuthal angle between two emitted hadrons in the final state of URHICs is quantified via the variance of its Fourier component $V_{2}$ over a large set of events~\cite{Ollitrault2023a}. In symmetric ultra-central collisions of present interest, this variance is proportional to the normalized variance of the quadrupole eccentricity $\langle \delta \epsilon^{2}_{2}   \rangle$ of the entropy density deposited in the transverse plane shortly after the reaction~\cite{Niemi:2016,Sousa:2024msh}. $\langle \delta \epsilon^{2}_{2}   \rangle$  can itself be related to the expectation value of the {\it squared eccentricity operator} $\mathcal{E}^{(2)}_{2}\equiv\mathcal{E}^{(1)}_{2}\mathcal{E}^{(1)}_{-2}$ in the ground-state of the collided nuclei~\cite{Duguet2025a} according to
\begin{align}
    \langle \delta \epsilon^{2}_{2}  \rangle &=  \lim_{s\to\infty} \langle \delta \epsilon^{2}_{2} \rangle(s) = \lim_{s\to\infty} \frac{1}{2} \frac{\langle \Psi^{0}_1(s) | \mathcal{E}^{(2)}_{2}| \Psi^{0}_1(s)\rangle }{\langle   \Psi^{0}_1(s)|  R^{(1)}_{2} | \Psi^{0}_1(s) \rangle^2} \label{meansquareeccentricity}    \, ,
\end{align}
where the one-body eccentricity and transverse radius operators respectively given by
\begin{equation}
 \mathcal{E}^{(1)}_{\pm2}(\mathbf{r}_{\perp}) \equiv r^2_{\perp} e^{\pm i2 \phi}  \,\,\, , \,\,\,  R^{(1)}_{2}(\mathbf{r}_{\perp}) \equiv r^2_{\perp} \, , \label{defoperators}
\end{equation}
depend on the nucleon position in the transverse plane defined by the coordinates $\mathbf{r}_{\perp} = (r_{\perp},\phi)$.

Analogously to the nuclear Hamiltonian, any operator has to be evolved via $U(s)$. In particular, the transformed squared eccentricity operator $\mathcal{E}^{(2)}_{2} (s)$ can be decomposed in terms of its zero-, one- and two-body components as 
\begin{equation}
\label{evolvedop}
U^{\dagger}(s) \mathcal{E}^{(2)}_{2}U(s) \equiv \mathcal{E}^{(2)\,(\text{0b})}_{2}(s)+\mathcal{E}^{(2)\,(\text{1b})}_{2}(s)+\mathcal{E}^{(2)\,(\text{2b})}_{2}(s)
\end{equation}
with $\mathcal{E}^{(2)\,(\text{0b})}_{2}(0) = 0$. Similar relation holds for $R^{(1)}_{2} (s)$ with $R^{(1)\,(\text{0b})}_{2}(0)=R^{(1)\,(\text{2b})}_{2}(0)=0$. Higher-rank components are discarded according to the IMSRG(2) approximation.

Considering the pure CR model and employing such a decomposition for $\mathcal{E}^{(2)}_{2} (s = 0)$ leads to a simple expression for the variance of the quadrupole eccentricity~\cite{Jia:2021tzt,Jia:2021qyu,Duguet2025a,Bofos:2026huw}
\begin{equation}
    \langle \delta \epsilon^2_2 \rangle^{\text{CR}} = \frac{1}{A} + \frac{3}{4\pi} \beta^2_{20} \, , \label{eccCRR}
\end{equation}
i.e., while the trivial one-body contribution originating from quantum fluctuations associated with the finite number of nucleons equates $A^{-1}$, the two-body contribution probing genuine two-body correlations scales with the square of the intrinsic quadrupole deformation parameter. Based on this relation, the dimensionless effective quadrupole-deformation~\cite{Blaizot2025a}
\begin{equation}
\label{eq:betaHE}
\mathcal{B}^2_{2}(\text{HE})
\equiv
\frac{4\pi}{3}
\langle
\delta \epsilon^{2,(2\text{b})}_2
\rangle ,
\end{equation}
is introduced to quantify quadrupolar two-body correlations. Although the notation is motivated by the CR result, $\mathcal{B}^2_{2}(\text{HE})$ is, in general, a signed measure of angular two-body quadrupole correlations rather than the literal square of an effective deformation parameter. It is therefore not required to be positive.

\begin{figure}[btp]     
  \centering             
  \includegraphics[width=0.7\columnwidth]{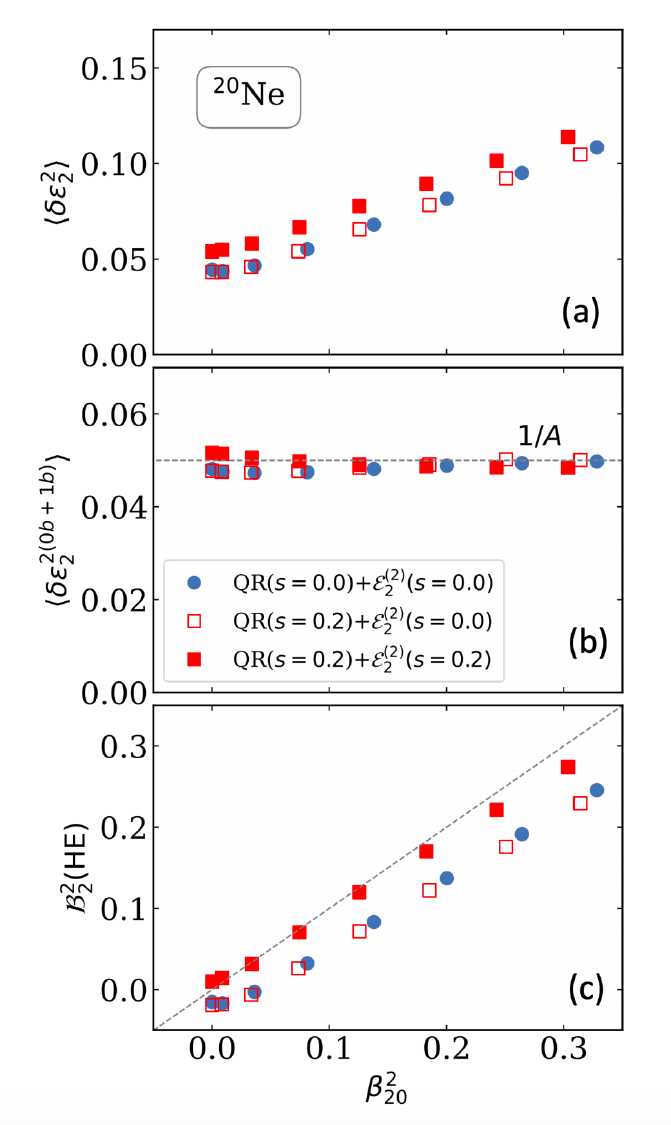}
  \caption{Results for the QR($s$) states of $^{20}\mathrm{Ne}$ at the flow parameter values $s=0$ and $s=0.2$, shown as a function of the constrained squared intrinsic quadrupole deformation $\beta_{20}^{2}(s)$. (a) Total normalized mean-square eccentricity $\langle\delta\epsilon_{2}^{2}\rangle$. (b) Sum of zero- and one-body contributions to $\langle\delta\epsilon_{2}^{2}\rangle$. (c) Squared effective quadrupole deformation $\mathcal{B}_{2}^{2}(\mathrm{HE})$. Filled blue circles (red squares) denote calculations at $s=0$ ($s=0.2)$ while open red squares denotes hybrid calculations, in which the evolved QR($s=0.2$) states are combined with unevolved operators ($s=0$) in the evaluation of $\langle\delta\epsilon_{2}^{2}\rangle$.  
  }
  \label{Fig1}
\end{figure}

\paragraph{Results and discussion.} 
Calculations of $^{16}$O and $^{20}$Ne are performed to analyze the contributions of distinct correlation mechanisms to the effective quadrupole deformation. A sHO basis characterized by $e_\mathrm{1max} = 6$ and $\hbar\omega = 16 $\,MeV is employed whereas the additional three-body truncation is set to $e_\mathrm{3max} = 14$. The convergence  of the results with respect to $e_\mathrm{1max}$ is demonstrated in the EM.

The top panel of Fig.~\ref{Fig1} displays the total normalized mean-square eccentricity $\langle\delta\epsilon^{2}_2\rangle$ in $^{20}\mathrm{Ne}$ in the QR($s$) approximation as a function of the squared intrinsic quadrupole deformation $\beta_{20}^2(s)$ for $s=0$ and $s=0.2$. While $\langle\delta\epsilon^{2}_2\rangle$ indeed scales with  $\beta_{20}^2$ as in the strict CR model, dynamical correlations are seen to systematically and non-negligibly increase it at fixed $\beta_{20}^2$~\footnote{The values of $\beta_{20}^2(s=0)$ and $\beta_{20}^2(s=0.2)$ differ slightly because the constraint is actually placed on the quadrupole moment $Q_{20}$ such that slightly different values of the mean-square radius at $s=0$ and $s=0.2$ lead to slightly  different values of the intrinsic deformation through Eq.~\eqref{defdef}.}. The middle panel proves that the combination of the zero- and one-body contributions (which reduces to the pure one-body term in the unevolved case) to $\langle \delta \epsilon^2_2 \rangle$ tracks $A^{-1}$ independently of $\beta_{20}^2$ and $s$ in agreement with Eq.~\ref{eccCRR}. Finally, two-body correlations imprinted into the two-body part of the mean-square eccentricity and quantified through the squared {\it effective} quadrupole deformation, as defined in Eq.~\eqref{eq:betaHE}, are shown in the bottom panel of Fig.~\ref{Fig1}. For unevolved calculations, $\mathcal{B}^2_{2}(\text{HE})$ is reduced compared to the CR approximation given by $\beta_{20}^2$ due to correlations induced by Pauli's exclusion principle~\cite{Bofos:2026nmg}. Including dynamical correlations through the IMSRG evolution ($s = 0.2$), however, happens to almost perfectly compensate for that effect. The hybrid results (empty squares) based on the QR($0.2$) states but the unevolved operators show that the content of the QR($0.2$) states is barely modified when constraining its intrinsic deformation to a fixed value. Therefore, the observed increase of $\mathcal{B}^2_{2}(\text{HE})$ is entirely driven by the explicit adjunction of dynamical correlations.

\begin{figure}[tbp]
  \centering
     \includegraphics[width=0.5\textwidth]{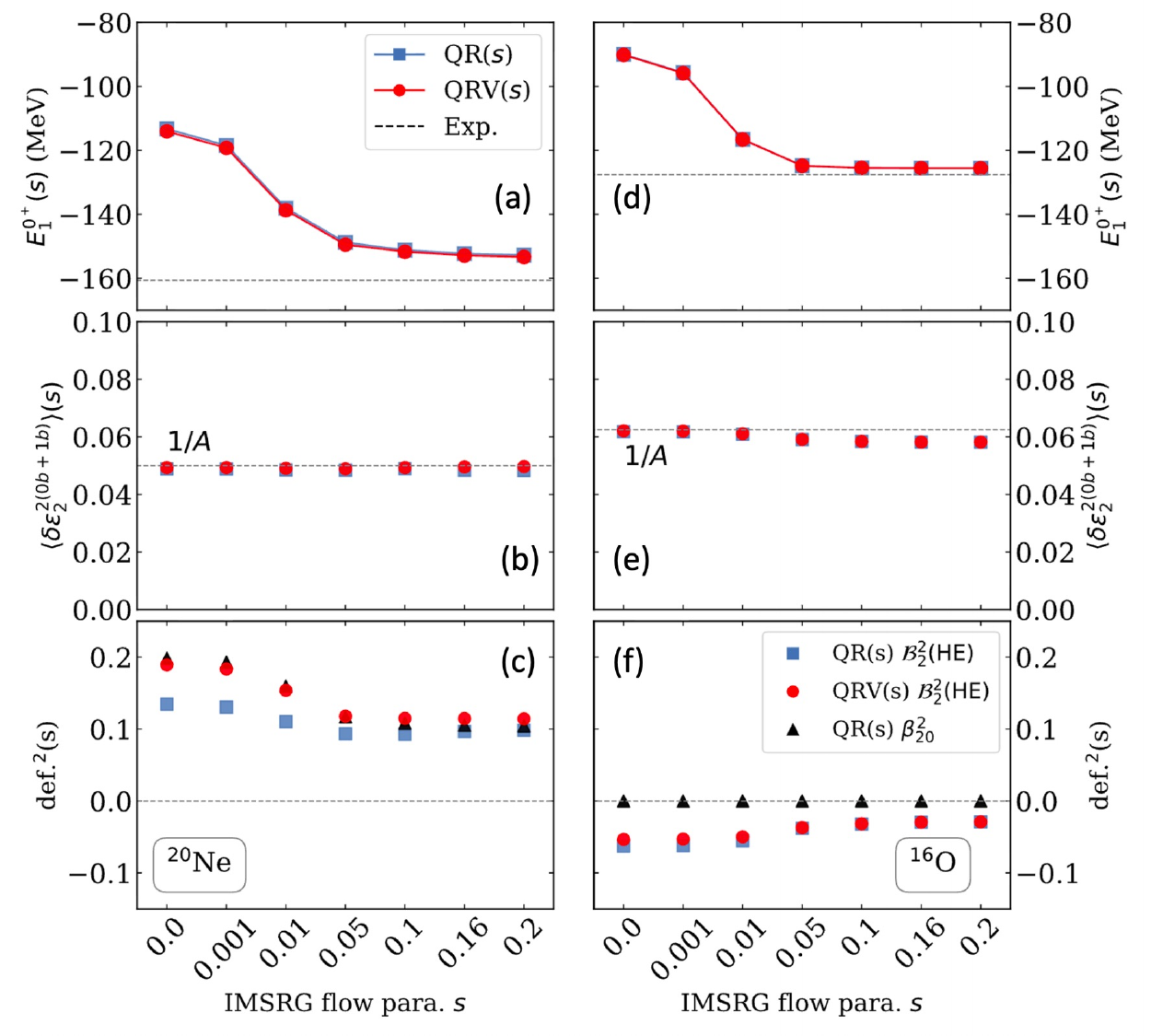}
  \caption{Results for the projected mean-field state QR($s$) of lowest energy and the final multi-reference state QRV($s$) for $^{20}$Ne (left) and $^{16}$O (right), shown as functions of the flow parameter $s$. Top panels (a, d): ground-state energies. Middle panels (b, e): sum of zero- and one-body contributions to the normalized mean-square eccentricity $\langle\delta\epsilon_{2}^{2}\rangle$. Bottom panels (c, f): squared effective  quadrupole deformation $\mathcal{B}_{2}^{2}(\mathrm{HE})$ and squared intrinsic quadrupole deformation $\beta_{20}^{2}$.}
  \label{Fig2}
\end{figure}

\begin{figure}[btp]     
  \centering             
  \includegraphics[width=\columnwidth]{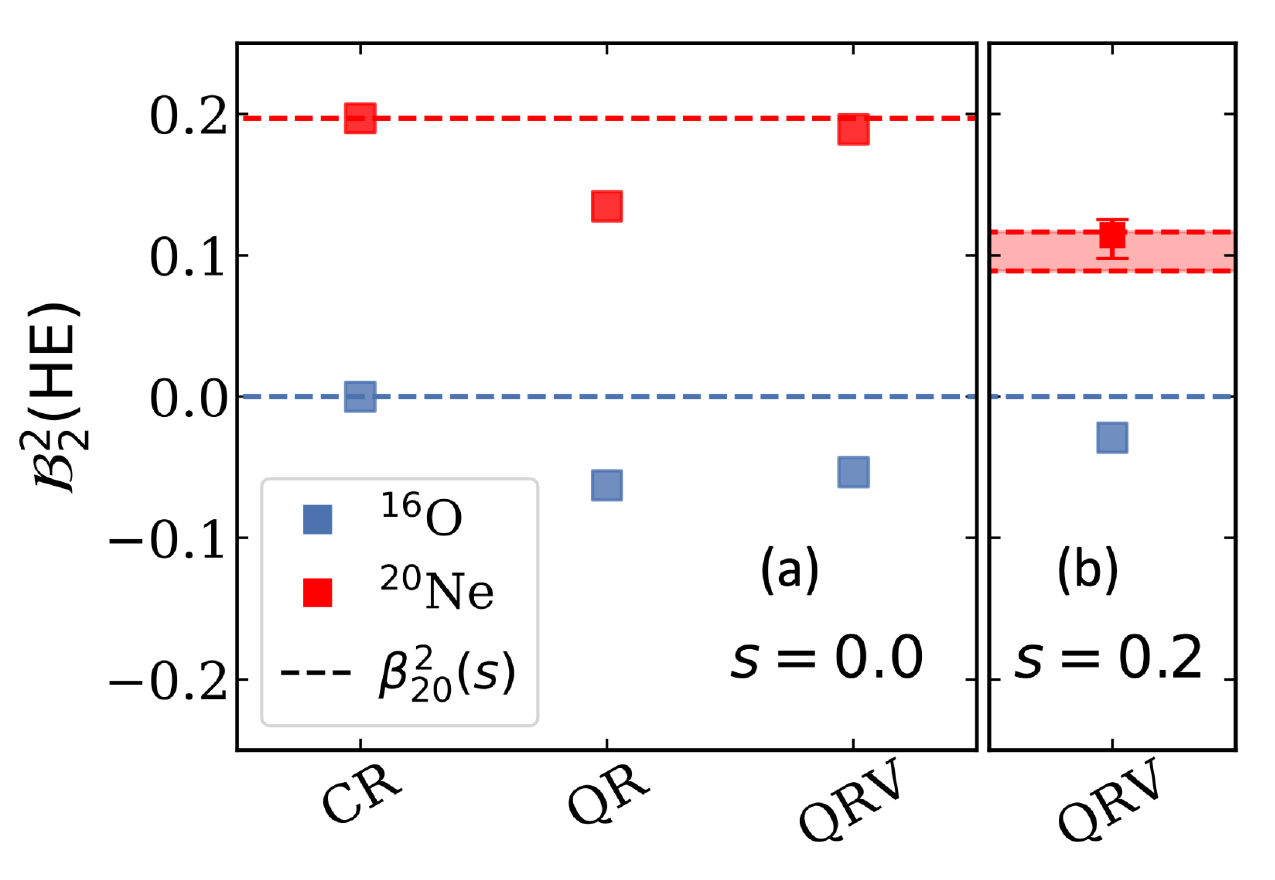}
  \caption{Squared effective quadrupole deformation $\mathcal{B}^2_{2}(\text{HE})$ in $^{20}\mathrm{Ne}$ and $^{16}\mathrm{O}$ as a function of the increasing many-body complexity/completeness. From left to right: CR($s=0$), QR($s=0$), QRV($s=0$) and QRV($s=0.2$). Horizontal lines denote the unevolved ($s=0$) and evolved ($s=0.2$) squared intrinsic quadrupole deformations. The results at $s=0.2$ are shown as a band (for $\beta^2_{20}$) and an error bar (for $\mathcal{B}_{2}^{2}(\mathrm{HE})$) to illustrate their dependence on the choice of ensemble reference state in the IMSRG calculation.  See the EM for details.}
  \label{Fig3}
\end{figure}

Figure~\ref{Fig2} displays results for the projected HFB state QR($s$) of lowest energy and the fully correlated state QRV($s$) as a function of $s$ in both  $^{20}\mathrm{Ne}$ (left) and $^{16}\mathrm{O}$ (right). The upper panel illustrates the energy gain brought by dynamical correlations, i.e., the ground-state energy is lowered by $39$ MeV ($34\%$) in $^{20}\mathrm{Ne}$ and  $35$ MeV ($40\%$) in $^{16}\mathrm{O}$. The agreement with experiment is thus drastically improved. Moreover, exploiting $e_\mathrm{1max}=4, 6$, and 8 calculations, the total energy of $^{20}$Ne extrapolated to the infinite basis-size limit is given by $-156.14$ MeV, deviating from the experimental value by 2.8\%, which is typical of IMSRG(2) calculations in light systems based on the EM1.8/2.0 Hamiltonian. Contrarily to dynamical correlations, shape fluctuations included into the QRV($s$) state have negligible impact on the ground-state energy. The middle panel confirms that the combined contribution of the zero- and one-body terms remains essentially unchanged and close to the $A^{-1}$ value.

Eventually the bottom panel illustrates the main findings of the present work. In absence of dynamical correlations ($s=0$), quadrupole shape fluctuations increase $\mathcal{B}^2_{2}(\text{HE})$ in $^{20}\mathrm{Ne}$ such that the negative contribution from Pauli's exclusion principle is almost perfectly compensated for. This makes $\mathcal{B}^2_{2}(\text{HE})$ be very close to the intrinsic deformation $\beta_{20}^2$ of the QR(0) state. Contrarily, quadrupole shape fluctuations have no influence in $^{16}\mathrm{O}$ such that $\mathcal{B}^2_{2}(\text{HE})$ remains essentially equal to the negative QR($s=0$) value entirely driven by the Pauli effect, consistent with the finding of Ref.\cite{Blaizot2025a}.  Adding dynamical correlations leads to an interesting combination of effects. In $^{20}\mathrm{Ne}$, the effective deformation $\mathcal{B}^2_{2}(\text{HE})$ is decreased by $42\%$, i.e., from $0.19$ down to $0.11$. This results from (i) the intrinsic deformation $\beta_{20}^2(s)$ of the QR($s$) state of lowest energy being itself decreased by about $50\%$, (ii) the positive contribution at fixed intrinsic deformation due to dynamical correlations compensating the impact of Pauli's exclusion principle and (iii) the effect from shape fluctuations being driven to zero due to an increased rigidity of the system as explicitly illustrated in the EM. Eventually, all three curves, i.e., the QR($s$)- and QRV($s$)-based $\mathcal{B}^2_{2}(\text{HE})$ as well as the QR($s$) intrinsic square deformation $\beta_{20}^2(s)$ converge to essentially the same value ($0.11$) at $s=0.2$. In $^{16}\mathrm{O}$, the intrinsic QR($s$) quadrupole deformation  $\beta_{20}^2(s)$ remains null for all $s$. The only impact of dynamical correlations is thus to increase $\mathcal{B}^2_{2}(\text{HE})$ by $40\%$ from $-0.05$ to $-0.03$ to almost compensate for correlations originating from Pauli's exclusion principle and approach the null intrinsic deformation. 

\paragraph{Summary.}  The present article provides the first detailed analysis of the various sources of two-body quadrupole correlations in $^{20}\mathrm{Ne}$ and $^{16}\mathrm{O}$ and how they differently impact the mean-square quadrupole eccentricity in ultra-central \oooo~and \nene~collisions. This so-called \emph{anatomy} of two-nucleon quadrupolar correlations relies in particular on the squared {\it effective} axial quadrupole deformation $\mathcal{B}^2_{2}(\text{HE})$ accessible in URHICs as well as on state-of-the-art \emph{ab initio} nuclear structure calculations combining the IMSRG and PGCM methods in a unified theoretical framework \cite{Yao:2020PRL,Frosini2022c,Duguet:2022zup,Zhou:2024vlt}.

Our main findings are summarized in Fig.~\ref{Fig3} for $^{20}\mathrm{Ne}$ and $^{16}\mathrm{O}$ where the values of $\mathcal{B}^2_{2}(\text{HE})$ is shown for four increasingly refined theoretical descriptions. While $\mathcal{B}^2_{2}(\text{HE})$ extracts exactly the squared intrinsic quadrupole deformation $\beta_{20}^2$ of the nucleus in the classical rotor approximation, it is shifted down by two-nucleon correlations associated with Pauli's exclusion principle in a quantum rotor improved description. Further incorporating collective correlations induced by quantum fluctuations of the intrinsic nuclear shape, the effect of Pauli's exclusion principle is almost perfectly compensated for (left untouched) in $^{20}\mathrm{Ne}$ ($^{16}\mathrm{O}$). Eventually, dynamical correlations associated with the non-collective nucleonic motions decrease (increase) $\mathcal{B}^2_{2}(\text{HE})$ by about $50\%$ in $^{20}\mathrm{Ne}$ ($^{16}\mathrm{O}$) such that  $\mathcal{B}^2_{2}(\text{HE})$ is eventually close to the value of $\beta_{20}^2$ of the nucleus {\it in presence} of such dynamical correlations. This result effectively justifies,  through an unexpected compensation process, the empirical success of analyses of ion-ion collisions based on an {\it effective} classical rotor picture~\cite{Duguet:2025qxi}.

\paragraph{Acknowledgment.}
We thank the organizers and participants of the YITP workshop \textit{``Intersection of Nuclear Structure and High-Energy Nuclear Collisions 2026''} for stimulating discussions, and Jiangyong Jia for carefully reading the manuscript and providing valuable comments. C.R.D., Y.L., and J.M.Y. are supported in part by the National Natural Science Foundation of China (Grant Nos.~125B2108, 12405143, and 12375119). This work has also received funding from the European Research Council under the European Union's Horizon Europe Research and Innovation Programme (Grant Agreement No.~101162059). The calculations were partly performed using computational resources provided by CCRT (TOPAZE supercomputer).


\begin{thebibliography}{42}%
\makeatletter
\providecommand \@ifxundefined [1]{%
 \@ifx{#1\undefined}
}%
\providecommand \@ifnum [1]{%
 \ifnum #1\expandafter \@firstoftwo
 \else \expandafter \@secondoftwo
 \fi
}%
\providecommand \@ifx [1]{%
 \ifx #1\expandafter \@firstoftwo
 \else \expandafter \@secondoftwo
 \fi
}%
\providecommand \natexlab [1]{#1}%
\providecommand \enquote  [1]{``#1''}%
\providecommand \bibnamefont  [1]{#1}%
\providecommand \bibfnamefont [1]{#1}%
\providecommand \citenamefont [1]{#1}%
\providecommand \href@noop [0]{\@secondoftwo}%
\providecommand \href [0]{\begingroup \@sanitize@url \@href}%
\providecommand \@href[1]{\@@startlink{#1}\@@href}%
\providecommand \@@href[1]{\endgroup#1\@@endlink}%
\providecommand \@sanitize@url [0]{\catcode `\\12\catcode `\$12\catcode
  `\&12\catcode `\#12\catcode `\^12\catcode `\_12\catcode `\%12\relax}%
\providecommand \@@startlink[1]{}%
\providecommand \@@endlink[0]{}%
\providecommand \url  [0]{\begingroup\@sanitize@url \@url }%
\providecommand \@url [1]{\endgroup\@href {#1}{\urlprefix }}%
\providecommand \urlprefix  [0]{URL }%
\providecommand \Eprint [0]{\href }%
\providecommand \doibase [0]{https://doi.org/}%
\providecommand \selectlanguage [0]{\@gobble}%
\providecommand \bibinfo  [0]{\@secondoftwo}%
\providecommand \bibfield  [0]{\@secondoftwo}%
\providecommand \translation [1]{[#1]}%
\providecommand \BibitemOpen [0]{}%
\providecommand \bibitemStop [0]{}%
\providecommand \bibitemNoStop [0]{.\EOS\space}%
\providecommand \EOS [0]{\spacefactor3000\relax}%
\providecommand \BibitemShut  [1]{\csname bibitem#1\endcsname}%
\let\auto@bib@innerbib\@empty
\bibitem [{\citenamefont {Giacalone}\ \emph {et~al.}(2018)\citenamefont
  {Giacalone}, \citenamefont {Noronha-Hostler}, \citenamefont {Luzum},\ and\
  \citenamefont {Ollitrault}}]{Giacalone2018a}%
  \BibitemOpen
  \bibfield  {author} {\bibinfo {author} {\bibfnamefont {G.}~\bibnamefont
  {Giacalone}}, \bibinfo {author} {\bibfnamefont {J.}~\bibnamefont
  {Noronha-Hostler}}, \bibinfo {author} {\bibfnamefont {M.}~\bibnamefont
  {Luzum}},\ and\ \bibinfo {author} {\bibfnamefont {J.-Y.}\ \bibnamefont
  {Ollitrault}},\ }\href {https://doi.org/10.1103/PhysRevC.97.034904}
  {\bibfield  {journal} {\bibinfo  {journal} {Phys. Rev. C}\ }\textbf {\bibinfo
  {volume} {97}},\ \bibinfo {pages} {034904} (\bibinfo {year}
  {2018})}\BibitemShut {NoStop}%
\bibitem [{\citenamefont {Giacalone}\ \emph {et~al.}(2021)\citenamefont
  {Giacalone}, \citenamefont {Jia},\ and\ \citenamefont
  {Som\`a}}]{Giacalone2021a}%
  \BibitemOpen
  \bibfield  {author} {\bibinfo {author} {\bibfnamefont {G.}~\bibnamefont
  {Giacalone}}, \bibinfo {author} {\bibfnamefont {J.}~\bibnamefont {Jia}},\
  and\ \bibinfo {author} {\bibfnamefont {V.}~\bibnamefont {Som\`a}},\ }\href
  {https://doi.org/10.1103/PhysRevC.104.L041903} {\bibfield  {journal}
  {\bibinfo  {journal} {Phys. Rev. C}\ }\textbf {\bibinfo {volume} {104}},\
  \bibinfo {pages} {L041903} (\bibinfo {year} {2021})}\BibitemShut {NoStop}%
\bibitem [{\citenamefont {Summerfield}\ \emph {et~al.}(2021)\citenamefont
  {Summerfield}, \citenamefont {Lu}, \citenamefont {Plumberg}, \citenamefont
  {Lee}, \citenamefont {Noronha-Hostler},\ and\ \citenamefont
  {Timmins}}]{Summerfield2021a}%
  \BibitemOpen
  \bibfield  {author} {\bibinfo {author} {\bibfnamefont {N.}~\bibnamefont
  {Summerfield}}, \bibinfo {author} {\bibfnamefont {B.-N.}\ \bibnamefont {Lu}},
  \bibinfo {author} {\bibfnamefont {C.}~\bibnamefont {Plumberg}}, \bibinfo
  {author} {\bibfnamefont {D.}~\bibnamefont {Lee}}, \bibinfo {author}
  {\bibfnamefont {J.}~\bibnamefont {Noronha-Hostler}},\ and\ \bibinfo {author}
  {\bibfnamefont {A.}~\bibnamefont {Timmins}},\ }\href
  {https://doi.org/10.1103/PhysRevC.104.L041901} {\bibfield  {journal}
  {\bibinfo  {journal} {Phys. Rev. C}\ }\textbf {\bibinfo {volume} {104}},\
  \bibinfo {pages} {L041901} (\bibinfo {year} {2021})}\BibitemShut {NoStop}%
\bibitem [{\citenamefont {Bally}\ \emph {et~al.}(2022)\citenamefont {Bally},
  \citenamefont {Bender}, \citenamefont {Giacalone},\ and\ \citenamefont
  {Som\`a}}]{Bally2022a}%
  \BibitemOpen
  \bibfield  {author} {\bibinfo {author} {\bibfnamefont {B.}~\bibnamefont
  {Bally}}, \bibinfo {author} {\bibfnamefont {M.}~\bibnamefont {Bender}},
  \bibinfo {author} {\bibfnamefont {G.}~\bibnamefont {Giacalone}},\ and\
  \bibinfo {author} {\bibfnamefont {V.}~\bibnamefont {Som\`a}},\ }\href
  {https://doi.org/10.1103/PhysRevLett.128.082301} {\bibfield  {journal}
  {\bibinfo  {journal} {Phys. Rev. Lett.}\ }\textbf {\bibinfo {volume} {128}},\
  \bibinfo {pages} {082301} (\bibinfo {year} {2022})}\BibitemShut {NoStop}%
\bibitem [{\citenamefont {Zhang}\ and\ \citenamefont {Jia}(2022)}]{Zhang2022a}%
  \BibitemOpen
  \bibfield  {author} {\bibinfo {author} {\bibfnamefont {C.}~\bibnamefont
  {Zhang}}\ and\ \bibinfo {author} {\bibfnamefont {J.}~\bibnamefont {Jia}},\
  }\href {https://doi.org/10.1103/PhysRevLett.128.022301} {\bibfield  {journal}
  {\bibinfo  {journal} {Phys. Rev. Lett.}\ }\textbf {\bibinfo {volume} {128}},\
  \bibinfo {pages} {022301} (\bibinfo {year} {2022})}\BibitemShut {NoStop}%
\bibitem [{\citenamefont {Ryssens}\ \emph {et~al.}(2023)\citenamefont
  {Ryssens}, \citenamefont {Giacalone}, \citenamefont {Schenke},\ and\
  \citenamefont {Shen}}]{Ryssens2023a}%
  \BibitemOpen
  \bibfield  {author} {\bibinfo {author} {\bibfnamefont {W.}~\bibnamefont
  {Ryssens}}, \bibinfo {author} {\bibfnamefont {G.}~\bibnamefont {Giacalone}},
  \bibinfo {author} {\bibfnamefont {B.}~\bibnamefont {Schenke}},\ and\ \bibinfo
  {author} {\bibfnamefont {C.}~\bibnamefont {Shen}},\ }\href
  {https://doi.org/10.1103/PhysRevLett.130.212302} {\bibfield  {journal}
  {\bibinfo  {journal} {Phys. Rev. Lett.}\ }\textbf {\bibinfo {volume} {130}},\
  \bibinfo {pages} {212302} (\bibinfo {year} {2023})}\BibitemShut {NoStop}%
\bibitem [{\citenamefont {{STAR Collaboration}}(2024)}]{STAR2024a}%
  \BibitemOpen
  \bibfield  {author} {\bibinfo {author} {\bibnamefont {{STAR
  Collaboration}}},\ }\href {https://doi.org/10.1038/s41586-024-08097-2}
  {\bibfield  {journal} {\bibinfo  {journal} {Nature}\ }\textbf {\bibinfo
  {volume} {635}},\ \bibinfo {pages} {67} (\bibinfo {year} {2024})}\BibitemShut
  {NoStop}%
\bibitem [{\citenamefont {Giacalone}\ \emph
  {et~al.}(2025{\natexlab{a}})\citenamefont {Giacalone}, \citenamefont {Bally},
  \citenamefont {Nijs}, \citenamefont {Shen}, \citenamefont {Duguet},
  \citenamefont {Ebran}, \citenamefont {Elhatisari}, \citenamefont {Frosini},
  \citenamefont {L\"ahde}, \citenamefont {Lee}, \citenamefont {Lu},
  \citenamefont {Ma}, \citenamefont {Mei\ss{}ner}, \citenamefont
  {Noronha-Hostler}, \citenamefont {Plumberg}, \citenamefont {Rodr\'{\i}guez},
  \citenamefont {Roth}, \citenamefont {van~der Schee},\ and\ \citenamefont
  {Som\`a}}]{Giacalone2025a}%
  \BibitemOpen
  \bibfield  {author} {\bibinfo {author} {\bibfnamefont {G.}~\bibnamefont
  {Giacalone}}, \bibinfo {author} {\bibfnamefont {B.}~\bibnamefont {Bally}},
  \bibinfo {author} {\bibfnamefont {G.}~\bibnamefont {Nijs}}, \bibinfo {author}
  {\bibfnamefont {S.}~\bibnamefont {Shen}}, \bibinfo {author} {\bibfnamefont
  {T.}~\bibnamefont {Duguet}}, \bibinfo {author} {\bibfnamefont {J.-P.}\
  \bibnamefont {Ebran}}, \bibinfo {author} {\bibfnamefont {S.}~\bibnamefont
  {Elhatisari}}, \bibinfo {author} {\bibfnamefont {M.}~\bibnamefont {Frosini}},
  \bibinfo {author} {\bibfnamefont {T.~A.}\ \bibnamefont {L\"ahde}}, \bibinfo
  {author} {\bibfnamefont {D.}~\bibnamefont {Lee}}, \bibinfo {author}
  {\bibfnamefont {B.-N.}\ \bibnamefont {Lu}}, \bibinfo {author} {\bibfnamefont
  {Y.-Z.}\ \bibnamefont {Ma}}, \bibinfo {author} {\bibfnamefont {U.-G.}\
  \bibnamefont {Mei\ss{}ner}}, \bibinfo {author} {\bibfnamefont
  {J.}~\bibnamefont {Noronha-Hostler}}, \bibinfo {author} {\bibfnamefont
  {C.}~\bibnamefont {Plumberg}}, \bibinfo {author} {\bibfnamefont {T.~R.}\
  \bibnamefont {Rodr\'{\i}guez}}, \bibinfo {author} {\bibfnamefont
  {R.}~\bibnamefont {Roth}}, \bibinfo {author} {\bibfnamefont {W.}~\bibnamefont
  {van~der Schee}},\ and\ \bibinfo {author} {\bibfnamefont {V.}~\bibnamefont
  {Som\`a}},\ }\href {https://doi.org/10.1103/k8rb-jgvq} {\bibfield  {journal}
  {\bibinfo  {journal} {Phys. Rev. Lett.}\ }\textbf {\bibinfo {volume} {135}},\
  \bibinfo {pages} {012302} (\bibinfo {year} {2025}{\natexlab{a}})}\BibitemShut
  {NoStop}%
\bibitem [{\citenamefont {Giacalone}\ \emph
  {et~al.}(2025{\natexlab{b}})\citenamefont {Giacalone}, \citenamefont {Zhao},
  \citenamefont {Bally}, \citenamefont {Shen}, \citenamefont {Duguet},
  \citenamefont {Ebran}, \citenamefont {Elhatisari}, \citenamefont {Frosini},
  \citenamefont {L\"ahde}, \citenamefont {Lee}, \citenamefont {Lu},
  \citenamefont {Ma}, \citenamefont {Mei\ss{}ner}, \citenamefont {Nijs},
  \citenamefont {Noronha-Hostler}, \citenamefont {Plumberg}, \citenamefont
  {Rodr\'{\i}guez}, \citenamefont {Roth}, \citenamefont {van~der Schee},
  \citenamefont {Schenke}, \citenamefont {Shen},\ and\ \citenamefont
  {Som\`a}}]{Giacalone2025b}%
  \BibitemOpen
  \bibfield  {author} {\bibinfo {author} {\bibfnamefont {G.}~\bibnamefont
  {Giacalone}}, \bibinfo {author} {\bibfnamefont {W.}~\bibnamefont {Zhao}},
  \bibinfo {author} {\bibfnamefont {B.}~\bibnamefont {Bally}}, \bibinfo
  {author} {\bibfnamefont {S.}~\bibnamefont {Shen}}, \bibinfo {author}
  {\bibfnamefont {T.}~\bibnamefont {Duguet}}, \bibinfo {author} {\bibfnamefont
  {J.-P.}\ \bibnamefont {Ebran}}, \bibinfo {author} {\bibfnamefont
  {S.}~\bibnamefont {Elhatisari}}, \bibinfo {author} {\bibfnamefont
  {M.}~\bibnamefont {Frosini}}, \bibinfo {author} {\bibfnamefont {T.~A.}\
  \bibnamefont {L\"ahde}}, \bibinfo {author} {\bibfnamefont {D.}~\bibnamefont
  {Lee}}, \bibinfo {author} {\bibfnamefont {B.-N.}\ \bibnamefont {Lu}},
  \bibinfo {author} {\bibfnamefont {Y.-Z.}\ \bibnamefont {Ma}}, \bibinfo
  {author} {\bibfnamefont {U.-G.}\ \bibnamefont {Mei\ss{}ner}}, \bibinfo
  {author} {\bibfnamefont {G.}~\bibnamefont {Nijs}}, \bibinfo {author}
  {\bibfnamefont {J.}~\bibnamefont {Noronha-Hostler}}, \bibinfo {author}
  {\bibfnamefont {C.}~\bibnamefont {Plumberg}}, \bibinfo {author}
  {\bibfnamefont {T.~R.}\ \bibnamefont {Rodr\'{\i}guez}}, \bibinfo {author}
  {\bibfnamefont {R.}~\bibnamefont {Roth}}, \bibinfo {author} {\bibfnamefont
  {W.}~\bibnamefont {van~der Schee}}, \bibinfo {author} {\bibfnamefont
  {B.}~\bibnamefont {Schenke}}, \bibinfo {author} {\bibfnamefont
  {C.}~\bibnamefont {Shen}},\ and\ \bibinfo {author} {\bibfnamefont
  {V.}~\bibnamefont {Som\`a}},\ }\href
  {https://doi.org/10.1103/PhysRevLett.134.082301} {\bibfield  {journal}
  {\bibinfo  {journal} {Phys. Rev. Lett.}\ }\textbf {\bibinfo {volume} {134}},\
  \bibinfo {pages} {082301} (\bibinfo {year} {2025}{\natexlab{b}})}\BibitemShut
  {NoStop}%
\bibitem [{\citenamefont {{STAR Collaboration}}(2025)}]{STAR:2025elk}%
  \BibitemOpen
  \bibfield  {author} {\bibinfo {author} {\bibnamefont {{STAR
  Collaboration}}},\ }\href {https://doi.org/10.1088/1361-6633/ae0fc3}
  {\bibfield  {journal} {\bibinfo  {journal} {Rept. Prog. Phys.}\ }\textbf
  {\bibinfo {volume} {88}},\ \bibinfo {pages} {108601} (\bibinfo {year}
  {2025})}\BibitemShut {NoStop}%
\bibitem [{\citenamefont {Niemi}\ \emph {et~al.}(2016)\citenamefont {Niemi},
  \citenamefont {Eskola},\ and\ \citenamefont {Paatelainen}}]{Niemi:2016}%
  \BibitemOpen
  \bibfield  {author} {\bibinfo {author} {\bibfnamefont {H.}~\bibnamefont
  {Niemi}}, \bibinfo {author} {\bibfnamefont {K.~J.}\ \bibnamefont {Eskola}},\
  and\ \bibinfo {author} {\bibfnamefont {R.}~\bibnamefont {Paatelainen}},\
  }\href {https://doi.org/10.1103/PhysRevC.93.024907} {\bibfield  {journal}
  {\bibinfo  {journal} {Phys. Rev. C}\ }\textbf {\bibinfo {volume} {93}},\
  \bibinfo {pages} {024907} (\bibinfo {year} {2016})}\BibitemShut {NoStop}%
\bibitem [{\citenamefont {Blaizot}\ and\ \citenamefont
  {Giacalone}(2025)}]{Blaizot:2025scr}%
  \BibitemOpen
  \bibfield  {author} {\bibinfo {author} {\bibfnamefont {J.-P.}\ \bibnamefont
  {Blaizot}}\ and\ \bibinfo {author} {\bibfnamefont {G.}~\bibnamefont
  {Giacalone}},\ }\href {https://doi.org/10.1140/epja/s10050-025-01679-2}
  {\bibfield  {journal} {\bibinfo  {journal} {Eur. Phys. J. A}\ }\textbf
  {\bibinfo {volume} {61}},\ \bibinfo {pages} {220} (\bibinfo {year} {2025})},\
  \Eprint {https://arxiv.org/abs/2504.15421} {arXiv:2504.15421 [nucl-th]}
  \BibitemShut {NoStop}%
\bibitem [{\citenamefont {Duguet}\ \emph
  {et~al.}(2025{\natexlab{a}})\citenamefont {Duguet}, \citenamefont
  {Giacalone}, \citenamefont {Jeon},\ and\ \citenamefont
  {Tichai}}]{Duguet2025a}%
  \BibitemOpen
  \bibfield  {author} {\bibinfo {author} {\bibfnamefont {T.}~\bibnamefont
  {Duguet}}, \bibinfo {author} {\bibfnamefont {G.}~\bibnamefont {Giacalone}},
  \bibinfo {author} {\bibfnamefont {S.}~\bibnamefont {Jeon}},\ and\ \bibinfo
  {author} {\bibfnamefont {A.}~\bibnamefont {Tichai}},\ }\href
  {https://doi.org/10.1103/v2z7-wlnr} {\bibfield  {journal} {\bibinfo
  {journal} {Phys. Rev. Lett.}\ }\textbf {\bibinfo {volume} {135}},\ \bibinfo
  {pages} {182301} (\bibinfo {year} {2025}{\natexlab{a}})}\BibitemShut
  {NoStop}%
\bibitem [{\citenamefont {Giacalone}\ \emph {et~al.}(2026)\citenamefont
  {Giacalone}, \citenamefont {Nijs},\ and\ \citenamefont {van~der
  Schee}}]{Giacalone:2026fat}%
  \BibitemOpen
  \bibfield  {author} {\bibinfo {author} {\bibfnamefont {G.}~\bibnamefont
  {Giacalone}}, \bibinfo {author} {\bibfnamefont {G.}~\bibnamefont {Nijs}},\
  and\ \bibinfo {author} {\bibfnamefont {W.}~\bibnamefont {van~der Schee}},\
  }\href@noop {} {\  (\bibinfo {year} {2026})},\ \Eprint
  {https://arxiv.org/abs/2606.03993} {arXiv:2606.03993 [nucl-th]} \BibitemShut
  {NoStop}%
\bibitem [{\citenamefont {Duguet}\ \emph
  {et~al.}(2025{\natexlab{b}})\citenamefont {Duguet}, \citenamefont
  {Giacalone}, \citenamefont {Som{\`a}},\ and\ \citenamefont
  {Zhou}}]{Duguet:2025qxi}%
  \BibitemOpen
  \bibfield  {author} {\bibinfo {author} {\bibfnamefont {T.}~\bibnamefont
  {Duguet}}, \bibinfo {author} {\bibfnamefont {G.}~\bibnamefont {Giacalone}},
  \bibinfo {author} {\bibfnamefont {V.}~\bibnamefont {Som{\`a}}},\ and\
  \bibinfo {author} {\bibfnamefont {Y.}~\bibnamefont {Zhou}},\ }\href
  {https://doi.org/10.1140/epja/s10050-025-01715-1} {\bibfield  {journal}
  {\bibinfo  {journal} {Eur. Phys. J. A}\ }\textbf {\bibinfo {volume} {61}},\
  \bibinfo {pages} {237} (\bibinfo {year} {2025}{\natexlab{b}})},\ \Eprint
  {https://arxiv.org/abs/2512.05874} {arXiv:2512.05874 [nucl-th]} \BibitemShut
  {NoStop}%
\bibitem [{Note1()}]{Note1}%
  \BibitemOpen
  \bibinfo {note} {The present description describes the anisotropic hadronic
  flow based on pure nucleonic degrees of freedom. While the impact of
  sub-nucleonic degrees of freedom is expected to be largely subleading, it has
  to be consistently evaluated in the future.}\BibitemShut {Stop}%
\bibitem [{\citenamefont {Bofos}\ \emph
  {et~al.}(2026{\natexlab{a}})\citenamefont {Bofos}, \citenamefont {Bally},
  \citenamefont {Duguet},\ and\ \citenamefont {Frosini}}]{Bofos:2026huw}%
  \BibitemOpen
  \bibfield  {author} {\bibinfo {author} {\bibfnamefont {S.}~\bibnamefont
  {Bofos}}, \bibinfo {author} {\bibfnamefont {B.}~\bibnamefont {Bally}},
  \bibinfo {author} {\bibfnamefont {T.}~\bibnamefont {Duguet}},\ and\ \bibinfo
  {author} {\bibfnamefont {M.}~\bibnamefont {Frosini}},\ }\href
  {https://doi.org/https://doi.org/10.1016/j.physletb.2026.140649} {\bibfield
  {journal} {\bibinfo  {journal} {Physics Letters B}\ ,\ \bibinfo {pages}
  {140649}} (\bibinfo {year} {2026}{\natexlab{a}})}\BibitemShut {NoStop}%
\bibitem [{\citenamefont {Bofos}\ \emph
  {et~al.}(2026{\natexlab{b}})\citenamefont {Bofos}, \citenamefont {Li},
  \citenamefont {Ding}, \citenamefont {Bally}, \citenamefont {Duguet},
  \citenamefont {Frosini},\ and\ \citenamefont {Yao}}]{Bofos:2026nmg}%
  \BibitemOpen
  \bibfield  {author} {\bibinfo {author} {\bibfnamefont {S.}~\bibnamefont
  {Bofos}}, \bibinfo {author} {\bibfnamefont {Y.}~\bibnamefont {Li}}, \bibinfo
  {author} {\bibfnamefont {C.}~\bibnamefont {Ding}}, \bibinfo {author}
  {\bibfnamefont {B.}~\bibnamefont {Bally}}, \bibinfo {author} {\bibfnamefont
  {T.}~\bibnamefont {Duguet}}, \bibinfo {author} {\bibfnamefont
  {M.}~\bibnamefont {Frosini}},\ and\ \bibinfo {author} {\bibfnamefont
  {J.}~\bibnamefont {Yao}},\ }\href@noop {} {\  (\bibinfo {year}
  {2026}{\natexlab{b}})},\ \Eprint {https://arxiv.org/abs/2605.28813}
  {arXiv:2605.28813 [nucl-th]} \BibitemShut {NoStop}%
\bibitem [{\citenamefont {Sousa}\ \emph {et~al.}(2024)\citenamefont {Sousa},
  \citenamefont {Noronha},\ and\ \citenamefont {Luzum}}]{Sousa:2024msh}%
  \BibitemOpen
  \bibfield  {author} {\bibinfo {author} {\bibfnamefont {J.}~\bibnamefont
  {Sousa}}, \bibinfo {author} {\bibfnamefont {J.}~\bibnamefont {Noronha}},\
  and\ \bibinfo {author} {\bibfnamefont {M.}~\bibnamefont {Luzum}},\ }\href
  {https://doi.org/10.1103/PhysRevC.110.044909} {\bibfield  {journal} {\bibinfo
   {journal} {Phys. Rev. C}\ }\textbf {\bibinfo {volume} {110}},\ \bibinfo
  {pages} {044909} (\bibinfo {year} {2024})}\BibitemShut {NoStop}%
\bibitem [{\citenamefont {{ATLAS Collaboration}}(2026)}]{ATLAS:2025nnt}%
  \BibitemOpen
  \bibfield  {author} {\bibinfo {author} {\bibnamefont {{ATLAS
  Collaboration}}},\ }\href {https://doi.org/10.1103/xqxz-8bhf} {\bibfield
  {journal} {\bibinfo  {journal} {Phys. Rev. C}\ }\textbf {\bibinfo {volume}
  {113}},\ \bibinfo {pages} {045205} (\bibinfo {year} {2026})},\ \Eprint
  {https://arxiv.org/abs/2509.05171} {arXiv:2509.05171 [nucl-ex]} \BibitemShut
  {NoStop}%
\bibitem [{\citenamefont {Hayrapetyan}\ \emph {et~al.}(2026)\citenamefont
  {Hayrapetyan} \emph {et~al.}}]{CMS:2025tga}%
  \BibitemOpen
  \bibfield  {author} {\bibinfo {author} {\bibfnamefont {A.}~\bibnamefont
  {Hayrapetyan}} \emph {et~al.} (\bibinfo {collaboration} {CMS}),\ }\href
  {https://doi.org/10.1103/26wx-tg6f} {\bibfield  {journal} {\bibinfo
  {journal} {Phys. Rev. Lett.}\ }\textbf {\bibinfo {volume} {137}},\ \bibinfo
  {pages} {082302} (\bibinfo {year} {2026})},\ \Eprint
  {https://arxiv.org/abs/2510.02580} {arXiv:2510.02580 [nucl-ex]} \BibitemShut
  {NoStop}%
\bibitem [{\citenamefont {Abualrob}\ \emph {et~al.}(2026)\citenamefont
  {Abualrob} \emph {et~al.}}]{ALICE:2025luc}%
  \BibitemOpen
  \bibfield  {author} {\bibinfo {author} {\bibfnamefont {I.~J.}\ \bibnamefont
  {Abualrob}} \emph {et~al.} (\bibinfo {collaboration} {ALICE}),\ }\href
  {https://doi.org/10.1103/gymp-vp87} {\bibfield  {journal} {\bibinfo
  {journal} {Phys. Rev. Lett.}\ }\textbf {\bibinfo {volume} {137}},\ \bibinfo
  {pages} {082301} (\bibinfo {year} {2026})},\ \Eprint
  {https://arxiv.org/abs/2509.06428} {arXiv:2509.06428 [nucl-ex]} \BibitemShut
  {NoStop}%
\bibitem [{\citenamefont {Constantin}\ \emph {et~al.}(2026)\citenamefont
  {Constantin}, \citenamefont {G{\"o}tz}, \citenamefont {Rosenkvist},\ and\
  \citenamefont {Elfner}}]{Constantin:2025ova}%
  \BibitemOpen
  \bibfield  {author} {\bibinfo {author} {\bibfnamefont {L.}~\bibnamefont
  {Constantin}}, \bibinfo {author} {\bibfnamefont {N.}~\bibnamefont
  {G{\"o}tz}}, \bibinfo {author} {\bibfnamefont {C.~B.}\ \bibnamefont
  {Rosenkvist}},\ and\ \bibinfo {author} {\bibfnamefont {H.}~\bibnamefont
  {Elfner}},\ }\href {https://doi.org/10.1103/nvq9-55gq} {\bibfield  {journal}
  {\bibinfo  {journal} {Phys. Rev. C}\ }\textbf {\bibinfo {volume} {113}},\
  \bibinfo {pages} {054901} (\bibinfo {year} {2026})},\ \Eprint
  {https://arxiv.org/abs/2509.05613} {arXiv:2509.05613 [nucl-th]} \BibitemShut
  {NoStop}%
\bibitem [{\citenamefont {Blaizot}\ \emph {et~al.}(2025)\citenamefont
  {Blaizot}, \citenamefont {Giacalone},\ and\ \citenamefont
  {Lovato}}]{Blaizot2025a}%
  \BibitemOpen
  \bibfield  {author} {\bibinfo {author} {\bibfnamefont {J.-P.}\ \bibnamefont
  {Blaizot}}, \bibinfo {author} {\bibfnamefont {G.}~\bibnamefont {Giacalone}},\
  and\ \bibinfo {author} {\bibfnamefont {A.}~\bibnamefont {Lovato}},\ }\href
  {https://arxiv.org/abs/2512.18926} {} (\bibinfo {year} {2025}),\ \Eprint
  {https://arxiv.org/abs/2512.18926} {arXiv:2512.18926 [nucl-th]} \BibitemShut
  {NoStop}%
\bibitem [{\citenamefont {Hammer}\ \emph {et~al.}(2020)\citenamefont {Hammer},
  \citenamefont {K\"onig},\ and\ \citenamefont {van Kolck}}]{Hammer2020a}%
  \BibitemOpen
  \bibfield  {author} {\bibinfo {author} {\bibfnamefont {H.-W.}\ \bibnamefont
  {Hammer}}, \bibinfo {author} {\bibfnamefont {S.}~\bibnamefont {K\"onig}},\
  and\ \bibinfo {author} {\bibfnamefont {U.}~\bibnamefont {van Kolck}},\ }\href
  {https://doi.org/10.1103/RevModPhys.92.025004} {\bibfield  {journal}
  {\bibinfo  {journal} {Rev. Mod. Phys.}\ }\textbf {\bibinfo {volume} {92}},\
  \bibinfo {pages} {025004} (\bibinfo {year} {2020})}\BibitemShut {NoStop}%
\bibitem [{\citenamefont {Epelbaum}\ \emph {et~al.}(2020)\citenamefont
  {Epelbaum}, \citenamefont {Krebs},\ and\ \citenamefont
  {Reinert}}]{Epelbaum2020a}%
  \BibitemOpen
  \bibfield  {author} {\bibinfo {author} {\bibfnamefont {E.}~\bibnamefont
  {Epelbaum}}, \bibinfo {author} {\bibfnamefont {H.}~\bibnamefont {Krebs}},\
  and\ \bibinfo {author} {\bibfnamefont {P.}~\bibnamefont {Reinert}},\ }\href
  {https://doi.org/10.3389/fphy.2020.00098} {\bibfield  {journal} {\bibinfo
  {journal} {Front. Phys.}\ }\textbf {\bibinfo {volume} {8}} (\bibinfo {year}
  {2020})}\BibitemShut {NoStop}%
\bibitem [{\citenamefont {Hebeler}\ \emph {et~al.}(2011)\citenamefont
  {Hebeler}, \citenamefont {Bogner}, \citenamefont {Furnstahl}, \citenamefont
  {Nogga},\ and\ \citenamefont {Schwenk}}]{Hebeler11a}%
  \BibitemOpen
  \bibfield  {author} {\bibinfo {author} {\bibfnamefont {K.}~\bibnamefont
  {Hebeler}}, \bibinfo {author} {\bibfnamefont {S.~K.}\ \bibnamefont {Bogner}},
  \bibinfo {author} {\bibfnamefont {R.~J.}\ \bibnamefont {Furnstahl}}, \bibinfo
  {author} {\bibfnamefont {A.}~\bibnamefont {Nogga}},\ and\ \bibinfo {author}
  {\bibfnamefont {A.}~\bibnamefont {Schwenk}},\ }\href
  {https://doi.org/10.1103/PhysRevC.83.031301} {\bibfield  {journal} {\bibinfo
  {journal} {Phys. Rev. C}\ }\textbf {\bibinfo {volume} {83}},\ \bibinfo
  {pages} {031301} (\bibinfo {year} {2011})}\BibitemShut {NoStop}%
\bibitem [{\citenamefont {Hergert}(2020)}]{Hergert20}%
  \BibitemOpen
  \bibfield  {author} {\bibinfo {author} {\bibfnamefont {H.}~\bibnamefont
  {Hergert}},\ }\href {https://doi.org/10.3389/fphy.2020.00379} {\bibfield
  {journal} {\bibinfo  {journal} {Front. in Phys.}\ }\textbf {\bibinfo {volume}
  {8}},\ \bibinfo {pages} {379} (\bibinfo {year} {2020})}\BibitemShut {NoStop}%
\bibitem [{\citenamefont {Hill}\ and\ \citenamefont
  {Wheeler}(1953)}]{Hill1953a}%
  \BibitemOpen
  \bibfield  {author} {\bibinfo {author} {\bibfnamefont {D.~L.}\ \bibnamefont
  {Hill}}\ and\ \bibinfo {author} {\bibfnamefont {J.~A.}\ \bibnamefont
  {Wheeler}},\ }\href {https://doi.org/10.1103/PhysRev.89.1102} {\bibfield
  {journal} {\bibinfo  {journal} {Phys. Rev.}\ }\textbf {\bibinfo {volume}
  {89}},\ \bibinfo {pages} {1102} (\bibinfo {year} {1953})}\BibitemShut
  {NoStop}%
\bibitem [{\citenamefont {Griffin}\ and\ \citenamefont
  {Wheeler}(1957)}]{Griffin1957a}%
  \BibitemOpen
  \bibfield  {author} {\bibinfo {author} {\bibfnamefont {J.~J.}\ \bibnamefont
  {Griffin}}\ and\ \bibinfo {author} {\bibfnamefont {J.~A.}\ \bibnamefont
  {Wheeler}},\ }\href {https://doi.org/10.1103/PhysRev.108.311} {\bibfield
  {journal} {\bibinfo  {journal} {Phys. Rev.}\ }\textbf {\bibinfo {volume}
  {108}},\ \bibinfo {pages} {311} (\bibinfo {year} {1957})}\BibitemShut
  {NoStop}%
\bibitem [{\citenamefont {Sheikh}\ \emph {et~al.}(2021)\citenamefont {Sheikh},
  \citenamefont {Dobaczewski}, \citenamefont {Ring}, \citenamefont {Robledo},\
  and\ \citenamefont {Yannouleas}}]{Sheikh21a}%
  \BibitemOpen
  \bibfield  {author} {\bibinfo {author} {\bibfnamefont {J.~A.}\ \bibnamefont
  {Sheikh}}, \bibinfo {author} {\bibfnamefont {J.}~\bibnamefont {Dobaczewski}},
  \bibinfo {author} {\bibfnamefont {P.}~\bibnamefont {Ring}}, \bibinfo {author}
  {\bibfnamefont {L.~M.}\ \bibnamefont {Robledo}},\ and\ \bibinfo {author}
  {\bibfnamefont {C.}~\bibnamefont {Yannouleas}},\ }\href
  {https://doi.org/10.1088/1361-6471/ac288a} {\bibfield  {journal} {\bibinfo
  {journal} {Journal of Physics G: Nuclear and Particle Physics}\ }\textbf
  {\bibinfo {volume} {48}},\ \bibinfo {pages} {123001} (\bibinfo {year}
  {2021})}\BibitemShut {NoStop}%
\bibitem [{\citenamefont {Bally}\ and\ \citenamefont
  {Bender}(2021)}]{Bally2021b}%
  \BibitemOpen
  \bibfield  {author} {\bibinfo {author} {\bibfnamefont {B.}~\bibnamefont
  {Bally}}\ and\ \bibinfo {author} {\bibfnamefont {M.}~\bibnamefont {Bender}},\
  }\href {https://doi.org/10.1103/PhysRevC.103.024315} {\bibfield  {journal}
  {\bibinfo  {journal} {Phys. Rev. C}\ }\textbf {\bibinfo {volume} {103}},\
  \bibinfo {pages} {024315} (\bibinfo {year} {2021})}\BibitemShut {NoStop}%
\bibitem [{\citenamefont {Hergert}\ \emph {et~al.}(2016)\citenamefont
  {Hergert}, \citenamefont {Bogner}, \citenamefont {Morris}, \citenamefont
  {Schwenk},\ and\ \citenamefont {Tsukiyama}}]{Hergert:2016PR}%
  \BibitemOpen
  \bibfield  {author} {\bibinfo {author} {\bibfnamefont {H.}~\bibnamefont
  {Hergert}}, \bibinfo {author} {\bibfnamefont {S.~K.}\ \bibnamefont {Bogner}},
  \bibinfo {author} {\bibfnamefont {T.~D.}\ \bibnamefont {Morris}}, \bibinfo
  {author} {\bibfnamefont {A.}~\bibnamefont {Schwenk}},\ and\ \bibinfo {author}
  {\bibfnamefont {K.}~\bibnamefont {Tsukiyama}},\ }\href
  {https://doi.org/10.1016/j.physrep.2015.12.007} {\bibfield  {journal}
  {\bibinfo  {journal} {Phys. Rept.}\ }\textbf {\bibinfo {volume} {621}},\
  \bibinfo {pages} {165} (\bibinfo {year} {2016})},\ \Eprint
  {https://arxiv.org/abs/1512.06956} {arXiv:1512.06956 [nucl-th]} \BibitemShut
  {NoStop}%
\bibitem [{\citenamefont {Hergert}(2017)}]{Hergert:2017PS}%
  \BibitemOpen
  \bibfield  {author} {\bibinfo {author} {\bibfnamefont {H.}~\bibnamefont
  {Hergert}},\ }\href {https://doi.org/10.1088/1402-4896/92/2/023002}
  {\bibfield  {journal} {\bibinfo  {journal} {Phys. Scripta}\ }\textbf
  {\bibinfo {volume} {92}},\ \bibinfo {pages} {023002} (\bibinfo {year}
  {2017})},\ \Eprint {https://arxiv.org/abs/1607.06882} {arXiv:1607.06882
  [nucl-th]} \BibitemShut {NoStop}%
\bibitem [{\citenamefont {Frosini}\ \emph {et~al.}(2022)\citenamefont
  {Frosini}, \citenamefont {Duguet}, \citenamefont {Ebran}, \citenamefont
  {Bally}, \citenamefont {Hergert}, \citenamefont {Rodr\'\i{}guez},
  \citenamefont {Roth}, \citenamefont {Yao},\ and\ \citenamefont
  {Som\`a}}]{Frosini2022c}%
  \BibitemOpen
  \bibfield  {author} {\bibinfo {author} {\bibfnamefont {M.}~\bibnamefont
  {Frosini}}, \bibinfo {author} {\bibfnamefont {T.}~\bibnamefont {Duguet}},
  \bibinfo {author} {\bibfnamefont {J.-P.}\ \bibnamefont {Ebran}}, \bibinfo
  {author} {\bibfnamefont {B.}~\bibnamefont {Bally}}, \bibinfo {author}
  {\bibfnamefont {H.}~\bibnamefont {Hergert}}, \bibinfo {author} {\bibfnamefont
  {T.~R.}\ \bibnamefont {Rodr\'\i{}guez}}, \bibinfo {author} {\bibfnamefont
  {R.}~\bibnamefont {Roth}}, \bibinfo {author} {\bibfnamefont {J.}~\bibnamefont
  {Yao}},\ and\ \bibinfo {author} {\bibfnamefont {V.}~\bibnamefont {Som\`a}},\
  }\href {https://doi.org/10.1140/epja/s10050-022-00694-x} {\bibfield
  {journal} {\bibinfo  {journal} {Eur. Phys. J. A}\ }\textbf {\bibinfo {volume}
  {58}},\ \bibinfo {pages} {64} (\bibinfo {year} {2022})},\ \Eprint
  {https://arxiv.org/abs/2111.01461} {arXiv:2111.01461 [nucl-th]} \BibitemShut
  {NoStop}%
\bibitem [{\citenamefont {Duguet}\ \emph {et~al.}(2023)\citenamefont {Duguet},
  \citenamefont {Ebran}, \citenamefont {Frosini}, \citenamefont {Hergert},\
  and\ \citenamefont {Som{\`a}}}]{Duguet:2022zup}%
  \BibitemOpen
  \bibfield  {author} {\bibinfo {author} {\bibfnamefont {T.}~\bibnamefont
  {Duguet}}, \bibinfo {author} {\bibfnamefont {J.~P.}\ \bibnamefont {Ebran}},
  \bibinfo {author} {\bibfnamefont {M.}~\bibnamefont {Frosini}}, \bibinfo
  {author} {\bibfnamefont {H.}~\bibnamefont {Hergert}},\ and\ \bibinfo {author}
  {\bibfnamefont {V.}~\bibnamefont {Som{\`a}}},\ }\href
  {https://doi.org/10.1140/epja/s10050-023-00914-y} {\bibfield  {journal}
  {\bibinfo  {journal} {Eur. Phys. J. A}\ }\textbf {\bibinfo {volume} {59}},\
  \bibinfo {pages} {13} (\bibinfo {year} {2023})},\ \Eprint
  {https://arxiv.org/abs/2209.03424} {arXiv:2209.03424 [nucl-th]} \BibitemShut
  {NoStop}%
\bibitem [{\citenamefont {Yao}\ \emph {et~al.}(2020)\citenamefont {Yao},
  \citenamefont {Bally}, \citenamefont {Engel}, \citenamefont {Wirth},
  \citenamefont {Rodr\'{\i}guez},\ and\ \citenamefont {Hergert}}]{Yao:2020PRL}%
  \BibitemOpen
  \bibfield  {author} {\bibinfo {author} {\bibfnamefont {J.~M.}\ \bibnamefont
  {Yao}}, \bibinfo {author} {\bibfnamefont {B.}~\bibnamefont {Bally}}, \bibinfo
  {author} {\bibfnamefont {J.}~\bibnamefont {Engel}}, \bibinfo {author}
  {\bibfnamefont {R.}~\bibnamefont {Wirth}}, \bibinfo {author} {\bibfnamefont
  {T.~R.}\ \bibnamefont {Rodr\'{\i}guez}},\ and\ \bibinfo {author}
  {\bibfnamefont {H.}~\bibnamefont {Hergert}},\ }\href
  {https://doi.org/10.1103/PhysRevLett.124.232501} {\bibfield  {journal}
  {\bibinfo  {journal} {Phys. Rev. Lett.}\ }\textbf {\bibinfo {volume} {124}},\
  \bibinfo {pages} {232501} (\bibinfo {year} {2020})}\BibitemShut {NoStop}%
\bibitem [{\citenamefont {Zhou}\ \emph {et~al.}(2025)\citenamefont {Zhou},
  \citenamefont {Ding}, \citenamefont {Yao}, \citenamefont {Bally},
  \citenamefont {Hergert}, \citenamefont {Jiao},\ and\ \citenamefont
  {Rodr{\'\i}guez}}]{Zhou:2024vlt}%
  \BibitemOpen
  \bibfield  {author} {\bibinfo {author} {\bibfnamefont {E.~F.}\ \bibnamefont
  {Zhou}}, \bibinfo {author} {\bibfnamefont {C.~R.}\ \bibnamefont {Ding}},
  \bibinfo {author} {\bibfnamefont {J.~M.}\ \bibnamefont {Yao}}, \bibinfo
  {author} {\bibfnamefont {B.}~\bibnamefont {Bally}}, \bibinfo {author}
  {\bibfnamefont {H.}~\bibnamefont {Hergert}}, \bibinfo {author} {\bibfnamefont
  {C.~F.}\ \bibnamefont {Jiao}},\ and\ \bibinfo {author} {\bibfnamefont
  {T.~R.}\ \bibnamefont {Rodr{\'\i}guez}},\ }\href
  {https://doi.org/10.1016/j.physletb.2025.139464} {\bibfield  {journal}
  {\bibinfo  {journal} {Phys. Lett. B}\ }\textbf {\bibinfo {volume} {865}},\
  \bibinfo {pages} {139464} (\bibinfo {year} {2025})},\ \Eprint
  {https://arxiv.org/abs/2410.23113} {arXiv:2410.23113 [nucl-th]} \BibitemShut
  {NoStop}%
\bibitem [{\citenamefont {Ollitrault}(2023)}]{Ollitrault2023a}%
  \BibitemOpen
  \bibfield  {author} {\bibinfo {author} {\bibfnamefont {J.-Y.}\ \bibnamefont
  {Ollitrault}},\ }\href {https://doi.org/10.1140/epja/s10050-023-01157-7}
  {\bibfield  {journal} {\bibinfo  {journal} {Eur. Phys. J. A}\ }\textbf
  {\bibinfo {volume} {59}},\ \bibinfo {pages} {236} (\bibinfo {year}
  {2023})}\BibitemShut {NoStop}%
\bibitem [{\citenamefont {Jia}(2022{\natexlab{a}})}]{Jia:2021tzt}%
  \BibitemOpen
  \bibfield  {author} {\bibinfo {author} {\bibfnamefont {J.}~\bibnamefont
  {Jia}},\ }\href {https://doi.org/10.1103/PhysRevC.105.014905} {\bibfield
  {journal} {\bibinfo  {journal} {Phys. Rev. C}\ }\textbf {\bibinfo {volume}
  {105}},\ \bibinfo {pages} {014905} (\bibinfo {year} {2022}{\natexlab{a}})},\
  \Eprint {https://arxiv.org/abs/2106.08768} {arXiv:2106.08768 [nucl-th]}
  \BibitemShut {NoStop}%
\bibitem [{\citenamefont {Jia}(2022{\natexlab{b}})}]{Jia:2021qyu}%
  \BibitemOpen
  \bibfield  {author} {\bibinfo {author} {\bibfnamefont {J.}~\bibnamefont
  {Jia}},\ }\href {https://doi.org/10.1103/PhysRevC.105.044905} {\bibfield
  {journal} {\bibinfo  {journal} {Phys. Rev. C}\ }\textbf {\bibinfo {volume}
  {105}},\ \bibinfo {pages} {044905} (\bibinfo {year}
  {2022}{\natexlab{b}})}\BibitemShut {NoStop}%
\bibitem [{Note2()}]{Note2}%
  \BibitemOpen
  \bibinfo {note} {The values of $\beta _{20}^2(s=0)$ and $\beta
  _{20}^2(s=0.2)$ differ slightly because the constraint is actually placed on
  the quadrupole moment $Q_{20}$ such that slightly different values of the
  mean-square radius at $s=0$ and $s=0.2$ lead to slightly different values of
  the intrinsic deformation through Eq.~\protect \eqref {defdef}.}\BibitemShut
  {Stop}%
\end{thebibliography}

%


\begin{center}
\textbf{End Matter}
\end{center}

\begin{figure}[htbp]
  \centering
  \includegraphics[width=0.8\columnwidth]{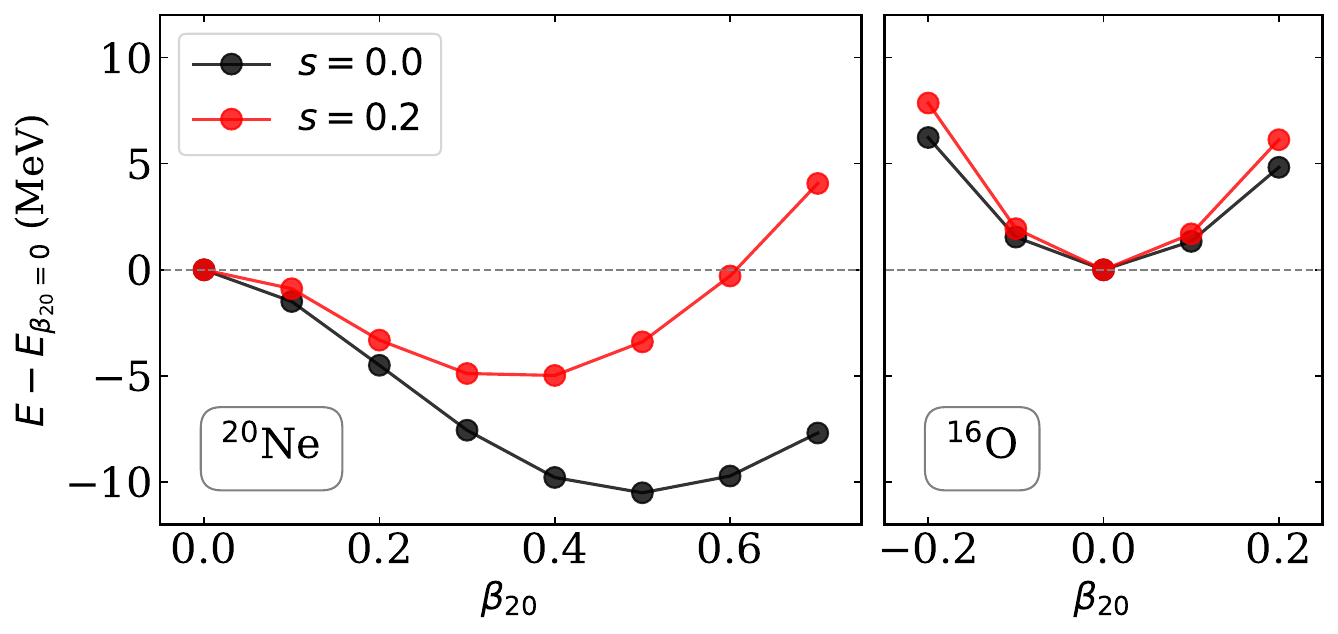}
  \caption{HFB mean-field TEC of $^{20}$Ne (left) and $^{16}$O (right) as a function of $\beta_{20}(s)$ for $s=0$ and $s=0.2$. For $^{20}$Ne, the reference state employed for the IMSRG evolution is taken as a statistical ensemble of the projected HFB (PHFB) states at the spherical and minimum points of the TEC computed for $s=0$ with mixing coefficients of 80\% and 20\%, respectively. For $^{16}$O, the reference state is taken as the sole PHFB state at the spherical point obtained for $s=0$. }
  \label{fig10}
\end{figure}

The HFB mean-field total energy curves (TECs) of $^{20}$Ne and $^{16}$O as functions of the intrinsic quadrupole deformation $\beta_{20}(s)$ are shown in Fig.~\ref{fig10} for $s=0$ and $s=0.2$. In $^{16}$O, the energy minimum remains spherical before and after the inclusion of dynamical correlations. In contrast, in $^{20}$Ne the equilibrium deformation decreases from $\beta_{20}(s=0)=0.5$ to $\beta_{20}(s=0.2)=0.35$ under the IMSRG evolution. Dynamical correlations also stiffen both nuclei against quadrupole deformation, as indicated by the increased curvature of the TEC near the minimum and the resulting suppression of ground-state shape fluctuations (see Fig.~\ref{Fig2}). Parabolic fits yield curvature increases from $113$ to $144$ MeV in $^{20}$Ne and from $275$ to $348$ MeV in $^{16}$O as $s$ evolves from $0$ to $0.2$. Although the relative increases are similar, the impact on shape fluctuations is more pronounced in $^{20}$Ne because the TEC of $^{16}$O is already stiff at $s=0$.

\begin{figure}[htbp]
  \centering
  \includegraphics[width=0.9\columnwidth]{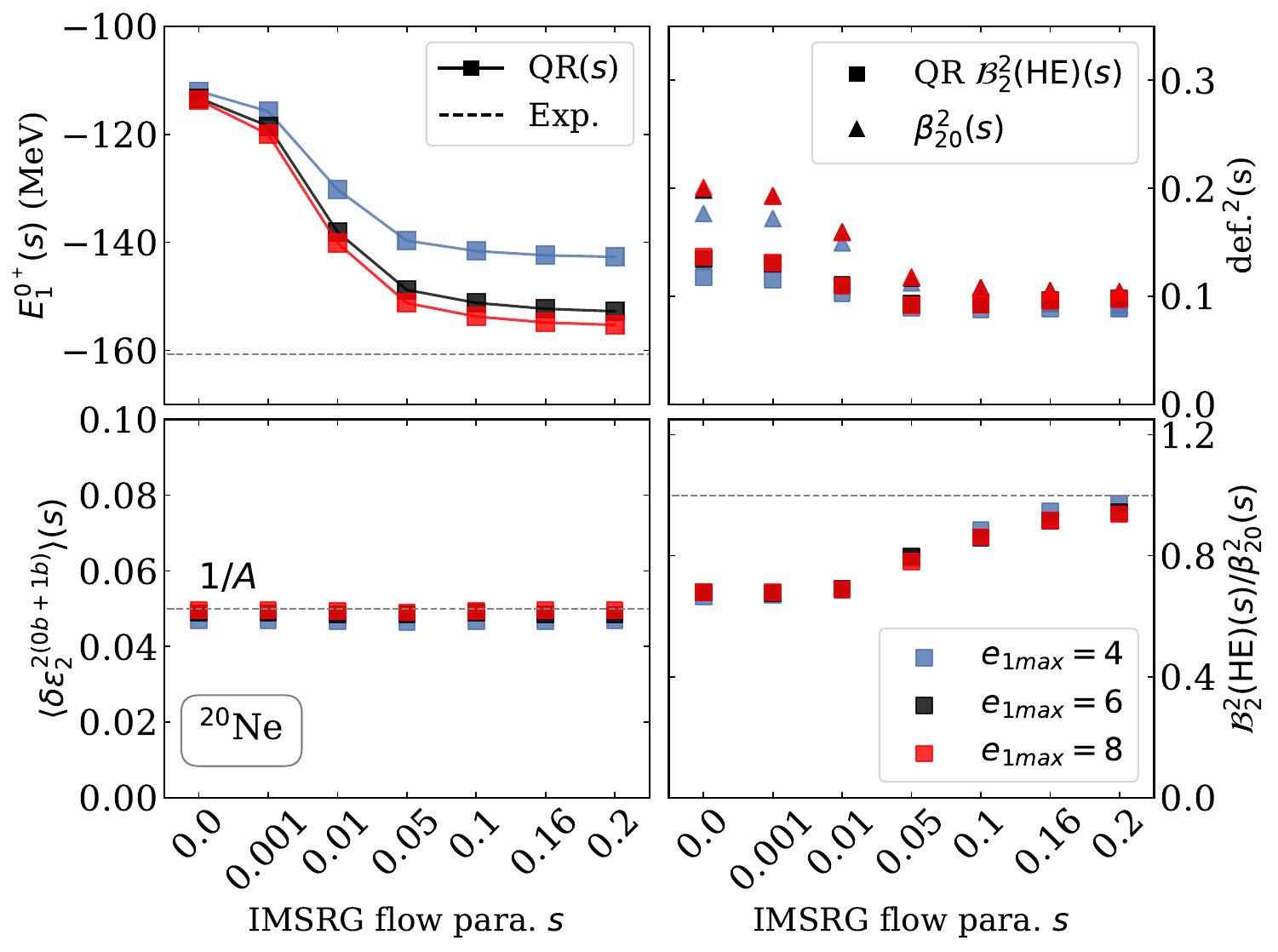}
  \caption{Convergence of the IMSRG+PHFB calculation results with respect to the model space $e_\mathrm{1max}=4,6,8$ in $^{20}$Ne as a function of the IMSRG flow parameter $s$. Left column and upper-right panel: same as fig.~\ref{Fig2}. Lower-right panel: ratio of the squared effective axial quadrupole deformation $\mathcal{B}^2_{2}(\text{HE})(s)$ to the squared intrinsic deformation $\beta_{20}^2(s)$. The reference state employed for the IMSRG evolution is taken as a statistical ensemble of the PHFB states at the spherical and minimum points of the TEC obtained for $s=0$ (see Fig.~\ref{fig10})  with mixing coefficients of 80\% and 20\%, respectively.}
  \label{fig8}
\end{figure} 

\begin{figure}[htbp]
  \centering
  \includegraphics[width=0.9\columnwidth]{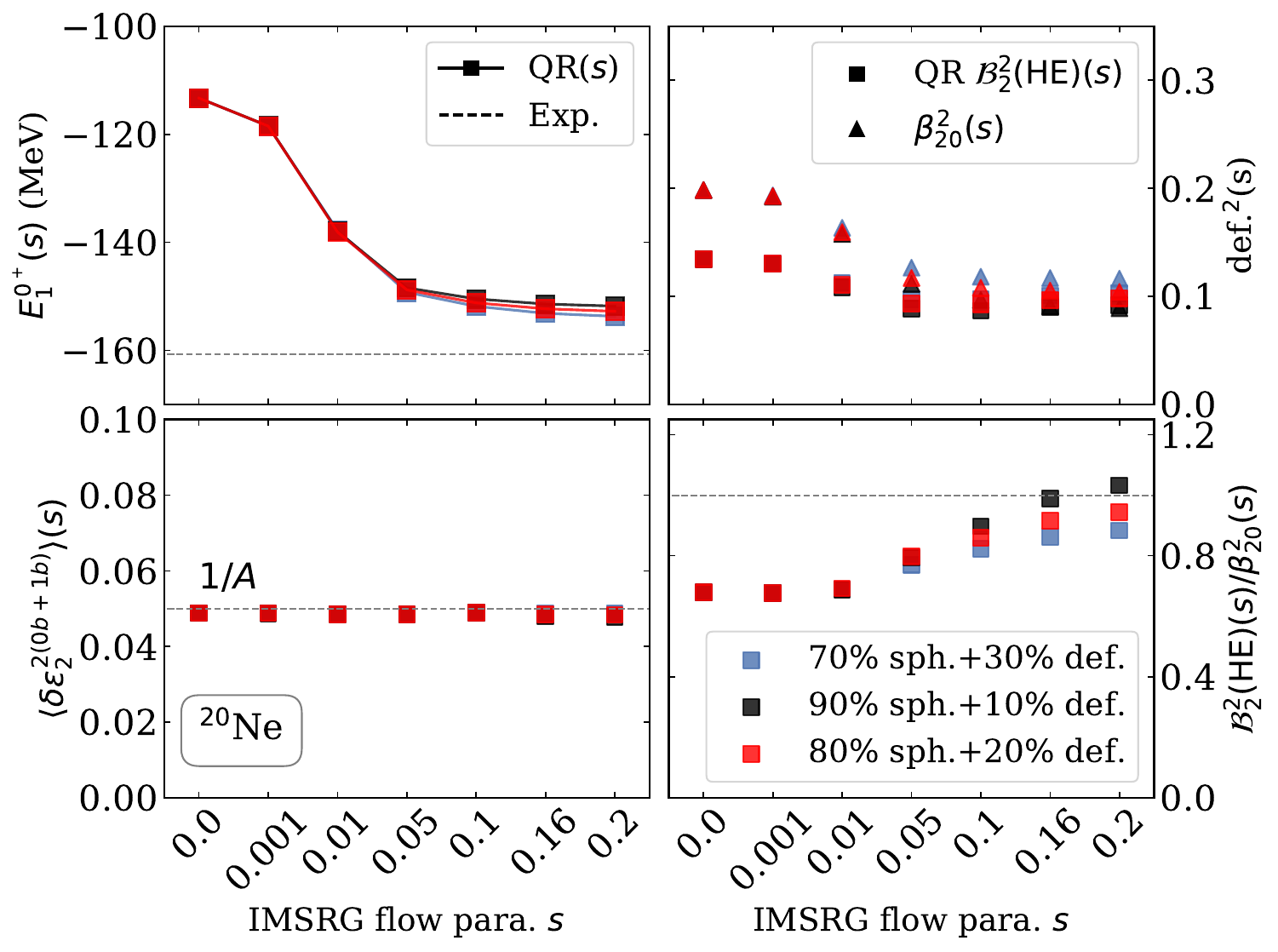}
  \caption{Same as Fig.~\ref{fig8} but for the convergence with respect to the mixing coefficients employed to define the ensemble reference state used for the IMSRG evolution. The results are obtained using the EM1.8/2.0 interaction (with 3NF) and $e_{\rm 1max}=6,\hbar\omega=16$ MeV.}
  \label{fig9}
\end{figure}

In order to check the convergence of the QR($s$) calculations presented in the main text, two additional figures are presently analyzed. Figure~\ref{fig8} investigates the convergence of the results obtained in $^{20}$Ne with respect to the one-body basis size characterized by  $e_\mathrm{1max}$.  The two left panels show, from top to bottom, the binding energy and the zero- plus one-body contributions to $\langle\delta\epsilon^2_2\rangle$, respectively. And the two right panels show the squared quadrupole deformation parameters and the ratio of $\mathcal{B}^2_{2}(\text{HE})$ to $\beta_{20}^2$.
The results are displayed for $e_\mathrm{1max}=4,6, 8$, respectively. For both the binding energy and the zero- plus one-body contributions to the normalized mean square eccentricity, the relative deviation between $e_\mathrm{1max}=6$ and $e_\mathrm{1max}=8$ results is less than $5\%$. The $e_\mathrm{1max}=6$ and $e_\mathrm{1max}=8$ squared deformation parameters are identical. This confirms that the $e_{\mathrm{1max}}=6$ model space adopted in this work is sufficiently large for the quantities considered in the present study.

Figure~\ref{fig9} further assesses the sensitivity of these three quantities to a $\pm 10\%$ variation in the mixing coefficients of the spherical and deformed PHFB configurations used to construct the IMSRG reference state. The variations are taken around the reference mixing weight adopted in this work. The resulting relative changes in the binding energy and in the zero- and one-body contributions to the normalized mean-square eccentricity remain below $5\%$. For the ratio $\mathcal{B}_{2}^{2}(\mathrm{HE})/\beta_{20}^{2}$, the sensitivity increases with the flow parameter $s$ but remains below $10\%$ at $s=0.2$.
As demonstrated in Figs.~\ref{fig8} and \ref{fig9}, all observables are sufficiently converged by $s=0.2$, supporting the termination of the IMSRG evolution at this flow-parameter value.

\end{document}